\RequirePackage{fix-cm}
\documentclass[twocolumn]{svjour3}          
\smartqed  
\usepackage{graphicx}

\usepackage{amsfonts}

\usepackage{amsmath}

\usepackage{amssymb}

\usepackage{mhchem} 

\usepackage{siunitx}

\newcommand{\slfrac}[2]{\left.#1\middle/#2\right.}

\usepackage{longtable}
\usepackage{bm}
\usepackage{multirow}
\usepackage{hyperref}
\usepackage{color}
\hypersetup{ colorlinks,
linkcolor=blue,
filecolor=black,
urlcolor=blue,
citecolor=blue }

\usepackage{subfigure}

\usepackage{sidecap}

 \usepackage{mathptmx}      
\begin{document}

\title{Modal interferometer sensor optimized for transverse misalignment}

\author{R. S. Grisotto  \and F. Beltr\'an-Mej\'ia \and P. F. Gomes
}

\institute{R. S. Grisotto  \and P. F. Gomes  \at
              Instituto de Ci\^encias Exatas e Tecnol\'{o}gicas, Universidade Federal de Goi\'{a}s, CEP 75801-615, Jata{\'i}, Brazil.
              \email{paulofreitasgomes@ufg.br}  
           \and
           F. Beltr\'an-Mej\'ia \at
              Instituto Nacional de Telecomunica\c{c}\~oes - INATEL, Santa Rita do Sapuca\'i - MG, Brazil.
}

\date{Received: date / Accepted: date}

\maketitle

\begin{abstract}
 We study the transmission coefficient and the transmitted power through an SMS sensor with transverse misalignment. We use the Finite Element Method to calculate the modes distribution by numerically solving the wave equation. The results show that the maximum transmission can be obtained when the misalignment is greater than zero due to the $n \neq m$ LP$_{nm}$ excited modes. Additionally, the transmitted power as function of the temperature shows that it is significant only in the aligned case.
\end{abstract}

\section{Introduction}

\begin{figure}[htbp]
\centering
\subfigure{a)
\includegraphics[width=1.8 in]{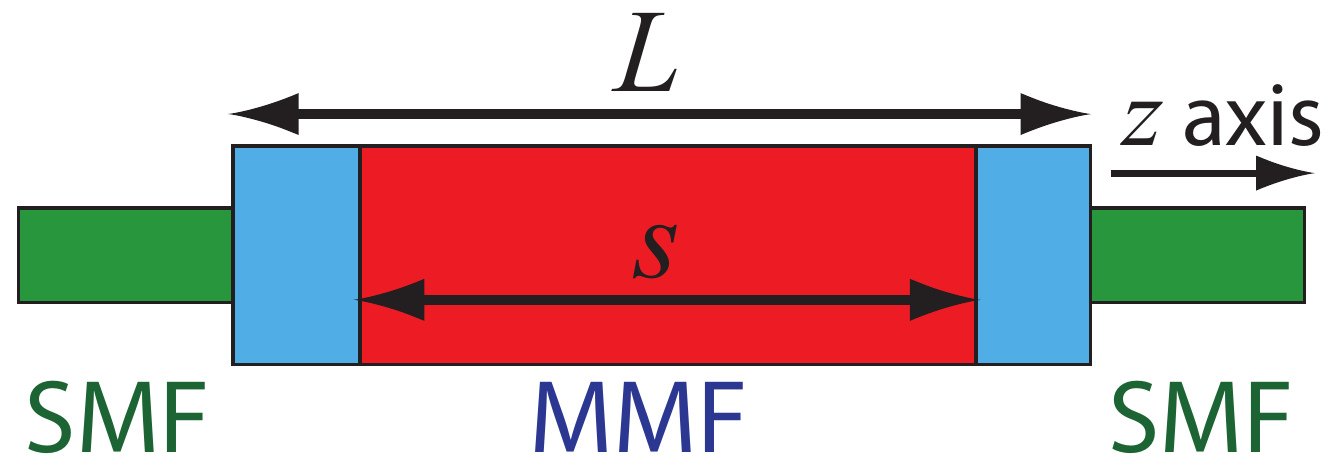}
\label{SMScopy}
}
\subfigure{b)
\includegraphics[width=1.5 in]{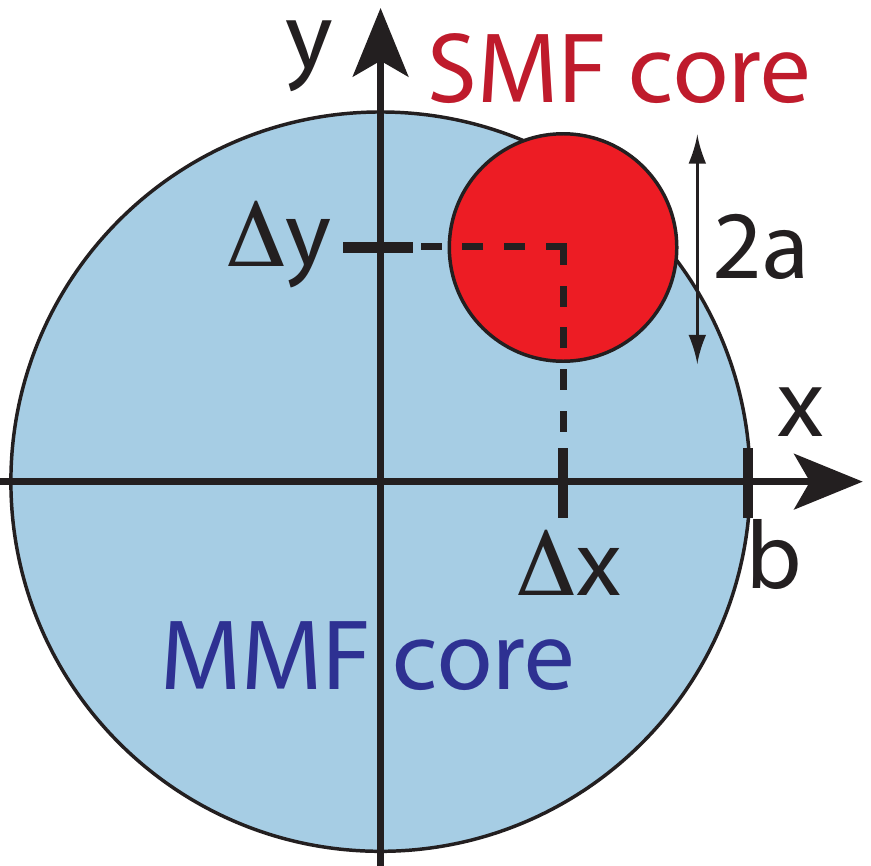}
\label{deltax}
}
\caption{\subref{SMScopy} Schematic diagram for the SMS structure. The SMF are the green ends, MMF is the blue/red region of length $L$ where the red part is the heated zone $s$. \subref{deltax} Illustration of the transverse misalignment between the two fibers in the $xy$ plane. The small red circle (radius $a$) is the SMF core and the large blue one (radius $b$) is the MMF core. The displacement between the two fibers is $\vec{d} = \hat{i} \Delta x + \hat{j} \Delta y$.}
\end{figure}

Fiber optic based sensors using intermodal interference have been intensely studied in basic research and in sensing applications such as strain and temperature \cite{Tripathi2009,Gao2010}, refractive index \cite{Wu2011,Zhao2014}, edge \cite{Hatta2009} and as a band-pass filter \cite{Tripathi2010}. The SMS (singlemode-multimode-singlemode) sensor is composed of a multimode fiber (MMF) spliced between two singlemode fibers (SMFs) \cite{Hatta2013,Kumar2014}, as illustrated on Fig. \ref{SMScopy}. In this device, the output transmission depend on external conditions such as temperature and strain. Another important parameter is the transverse misalignment between the longitudinal axis ($z$) of the fibers in the two splicing interfaces. In most cases, a perfect alignment --no transverse displacement in the $xy$ plane between the fiber ends-- is wanted. However, in practice there is always a small lateral misalignment due to the experimental errors caused by fusion splicing or free-space misalignment, and in some cases this displacement can be useful for sensing applications \cite{Dong2014}. 

Experimental techniques have been developed to measure the transmission after the first interface \cite{Flamm2012,Kaiser2009,Schulze2013} enabling the alignment by measuring the displacement with a microscope or maximizing the transmission. Flamm \textit{et al} \cite{Flamm2013} measured the transmission of different modes from one SMF to a MMF and observed that for circular symmetric modes LP$_{01}$ and LP$_{02}$ the transmission have its maxima when the two fibers are perfectly aligned. For not circular modes LP$_{11}$ and LP$_{21}$ the maxima are for a non zero displacement. In a more realistic scenario, when the three fibers are not axially aligned (lateral displacement greater than zero), LP$_{nm}$ modes without circular symmetry ($n \neq m$) will be excited in the MMF \cite{Flamm2013}. 

 Here, we study the transmission coefficient and the transmitted power transmitted through a SMS sensor as function of both an arbitrary transverse misalignment and the temperature. In the next section we introduce the theoretical concepts needed for the modal analysis. Later, in the results, we used the Finite Element Method (FEM) to calculate the mode distribution by numerically solving the full-vectorial wave equation. A coupling analysis is presented for the various modes excited in the MMF fiber. Finally, results and discussion lead towards the concluding remarks.

\section{Model} \label{sec:model}

We chose a germanium doped multimode fiber and calculate its material dispersion using the Mixed Sellmeier equation \cite{Fleming10984},
\begin{equation}
n(x,\lambda) = \sqrt{1+\dfrac{A_1 \lambda^2}{\lambda^2 - l_1^2} + \dfrac{A_2 \lambda^2}{\lambda^2 - l_2^2} + \dfrac{A_3 \lambda^2}{\lambda^2 - l_3^2}}  \label{sellmeiertrestermos},
\end{equation}
where $A_i=A_i(x)$ (the oscillator strength) and $l_i=l_i(x)$ (the oscillator wavelength) depend on the Ge concentration $x$ (the values of these constants are in Ref. \cite{Fleming10984}). We assumed linear interpolation between the two materials to calculate the $A_i$ and $l_i$ as function of $x$.

The single mode fiber used in all the simulations has core radius $a$ with refraction index $n_1$ and $n_0$ for the core and cladding layer, respectively. At \SI{24}{\celsius} (room temperature), the MMF has core radius $b$ and refraction index $n(x_g,\lambda_0)$ and $n(x_s,\lambda_0)$ for the core and cladding layer, as in~\eqref{sellmeiertrestermos}. The concentrations of germanium for the core and the cladding layer are $x_g$ and $x_s$, respectively. The radius for the cladding layer on both fibers is $c$. The values for all these parameters are on Table\,\ref{parametros}.
\begin{table}[h!]
\centering
\begin{tabular}{ccc}
\hline \hline
Symbol & Value & Description   \\
\hline
\hline $n_0$ & $1.4397$ & refractive index   \\  
\hline $n_1$ & $1.444$ &  refractive index  \\  
\hline $\lambda_0$ & \SI{1.55}{\micro\meter} & wavelength   \\  
\hline $a$ & \SI{4}{\micro\meter} & SMF core radius \\
\hline $b$ & \SI{30}{\micro\meter} & MMF core radius \\
\hline $c$ & \SI{60}{\micro\meter} & cladding radius \\
\hline $x_g$ & 19.34\,\% & Ge mole fraction for the MMF \\
\hline $x_s$ & 3.3\,\% & Ge mole fraction for the SMF \\
\hline
\hline
\end{tabular}
\caption{Parameters used to model the SMS sensor.}
\label{parametros}
\end{table}

\subsection{Transmitted power with transverse misalignment}

In our model a SMF couples light into a transverse misaligned MMF by an amount $\Delta x$ and $\Delta y$ in the $x$ and $y$ directions as shown in Fig.~\ref{deltax}. The electric field of the optical modes on the fiber were computed from the wave equation: $\vec{\nabla} \times  ( \vec{\nabla} \times  \vec{E} ) - k_0^2n^2 \vec{E} = 0$ with $k_0 = (2\pi)/\lambda_0$. We label $E_1$ as the single mode on SMF, $E_2$ as the fundamental mode on MMF and $E_3$, $E_4$, $E_5$ and $E_6$ as the excited higher order modes on the MMF. At the first interface $E_1$ can be represented as an expansion of the MMF modes,
\begin{equation}
E_1(x,y) = \Psi(x,y,0)= \sum_{i=2}^6 T_i E_i(x,y) \nonumber.
\end{equation}
Because we are considering non circular symmetric modes, we wrote the mode spatial dependence as $x,y$ instead of $r = \sqrt{x^2+y^2}$. The propagated mode at MMF at a distance $z$ from the first interface is 
\begin{equation}
\Psi(x,y,z) = \sum_{i=2}^6 T_i E_i(x,y) \exp \left( j \beta_i z\right)  \label{asdfqeerttyyy},
\end{equation}
where $j = \sqrt{-1}$ and $\beta_i$ is the propagation constant 
of the mode $E_i$. The coefficients $T_i$ are the transmission intensities through one interface considering different modes on the MMF. In our case, we computed them as function of the transverse misalignment,
\begin{equation}
T_i (\Delta x, \Delta y) = \dfrac{S_i^2}{N_1N_i}  \label{transmisaofundamental}. 
\end{equation}
The normalization integral is $N_i = \iint \vert E_i \vert^2 dxdy$ and the superposition one is $S_i = \iint  E_1 E_i  dxdy$ with $i=$ 2, 3, 4, 5 and 6. For example, $T_3$ is the transmission intensity for the $E_3$ mode on the MMF. We computed $T_i$ for the displacements $-b<\Delta x,\Delta y<b$, by numerically evaluating the integrals over all the computational domain \cite{Kiusalaas2005}. The total transmission $T_t$ is the sum of all the transmissions coefficients,
\begin{equation}
T_t = \sum_{i=2}^6 T_i \label{transmissaototal}.
\end{equation}
All the calculations were numerically implemented on the open source platform R \cite{plataformaR}. The reflection of the incident wave in the first interface between SMF and MMF (due to the difference of the refraction indexes) was neglected \cite{Wang2008}.

\subsection{Fractional modal power of the SMS device}

The transmission through the SMS sensor involves the transmission coefficient by two interfaces with an interference effect on the second one. Using~\eqref{asdfqeerttyyy}, the transmitted power of the SMS device considering a length $L$ is \cite{Tripathi2009,Mejia},
\begin{equation}
P_t = \left\vert \sum_{i=2}^6  T_i^2 \exp \left[ jL \left(  \beta_2 - \beta_i\right) \right] \right\vert^2    \label{powertotal}   .
\end{equation}
In order to better verify the contribution of the first and second modes ($E_2$ and $E_3$) we also computed their power,
\begin{equation}
P_{23} = \left\vert T_2^2 + T_3^2 \exp \left[ jL \left(  \beta_2 - \beta_3\right) \right]  \right\vert^2   \label{powerP23} .
\end{equation}
The transmission loss \cite{Gao2012} is defined as,
\begin{equation}
L_t = 10 \log (P_t) \qquad \therefore \qquad L_{23} = 10 \log (P_{23}) \label{dfgbnmzaqwe}.
\end{equation}

In order to illustrate the practical implementation of the misaligned SMS, we studied the effect of the temperature $t$ on the computed powers by using the temperature dependence of the MMF refraction index \cite{Tripathi2009},
\begin{equation}
n (t)= n_0+ \frac{dn}{dt} (t-t_0) ,
\end{equation}
where $n_0$ represents the refractive index at room temperature $t_0 =24^{\circ}C$ and $\slfrac{dn}{dt}$ is the thermo-optic coefficient. The values of $\slfrac{dn}{dt}$ for \ce{SiO_2} glass with different concentrations of \ce{GeO_2} are on Table~\ref{thermooptic}, assuming linear dependence on Ge concentration. 

\begin{table}[htbp]
\centering
\begin{tabular}{c|c|c|c|c}
\hline
$x$ (\%) & 0 & 15 & $x_g$ & $x_s$    \\
\hline $\slfrac{dn}{dt}$ ($\slfrac{10^{-5}}{\si{\celsius}})$  & 1.06  \cite{Kim2002} & 1.24 \cite{Kim2002}  &  1.29 & 1.1  \\ 
\hline
\end{tabular}
\caption{Values for the thermo-optic coefficient $\slfrac{dn}{dt}$ for a fiber with different concentrations of \ce{GeO_2}.}
\label{thermooptic}
\end{table}

We consider only a small segment $s<L$ of the MMF to be heated (see Fig. \ref{SMScopy}). The length $s$ and radius $b$ of the MMF at temperature $t$ are $s(t) = s_0 + \alpha s_0 ( t-t_0)$ and $b(t) = b_0+ \alpha b_0 ( t-t_0)$
where $s_0$ and $b_0$ are the values at temperature $t_0$. We used $\alpha = 5\times 10^{-5}/^{\circ}C$, which is the value for fused silica \cite{Huang1990}. The transmission coefficients $T_i$ will be affected by temperature due to the refraction index $n(t)$ and the core radius $b(t)$. The temperature dependence of the phase is written as \cite{Tripathi2009}:
\begin{equation}
\varphi_i (t) =  \left[ \beta_2(t_0) - \beta_i (t_0) \right] (L-s_0) +  \left[ \beta_2(t) - \beta_i(t) \right]  s(t)  \nonumber
\end{equation} 
where $\beta_i(t)$ is the temperature-dependent propagation constant of the $i$th mode. 

\begin{figure}[h!]
\centering
\subfigure{a)
\includegraphics[width=1.4 in]{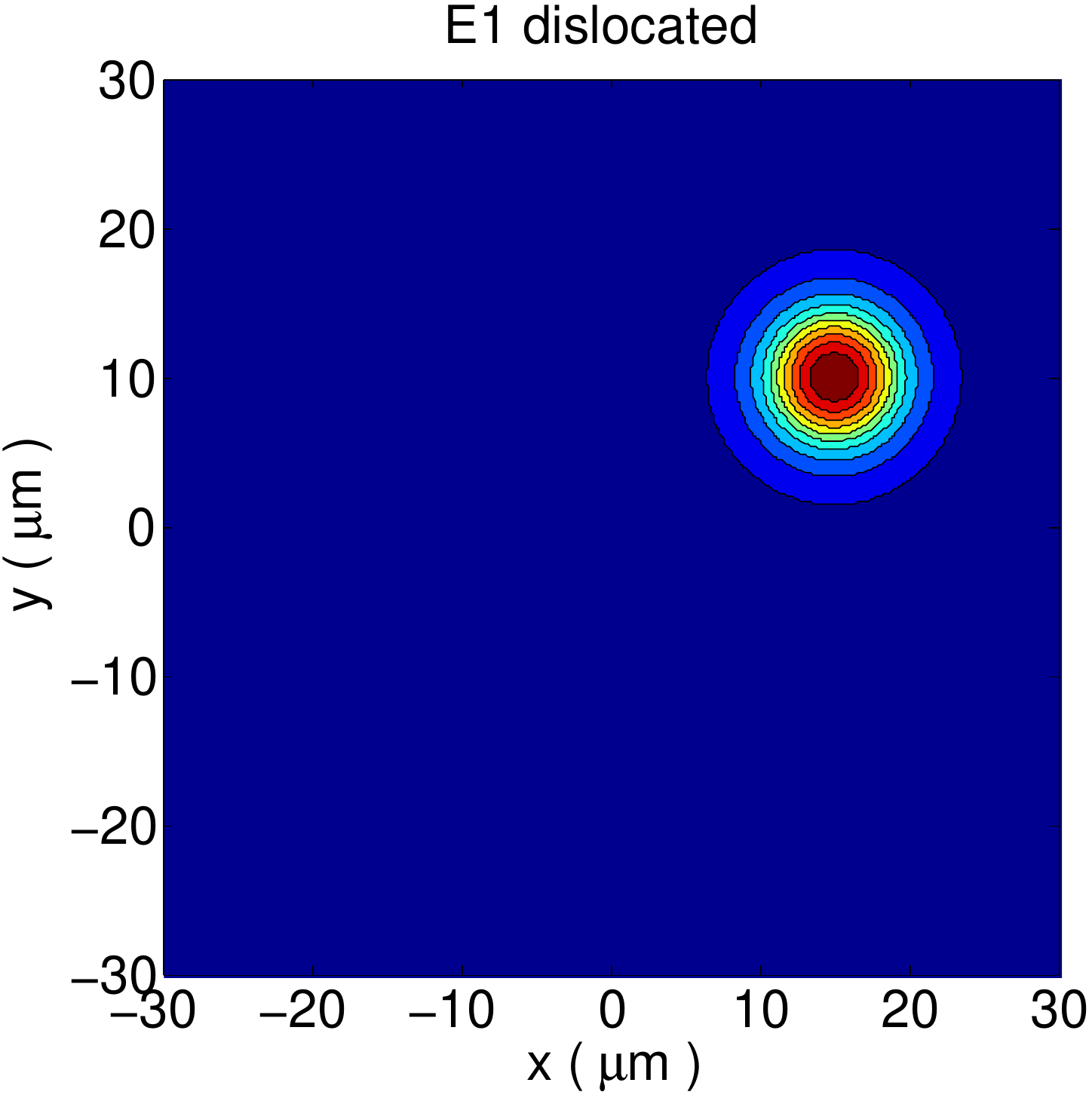}
\label{E1}
}
\subfigure{b)
\includegraphics[width=1.4 in]{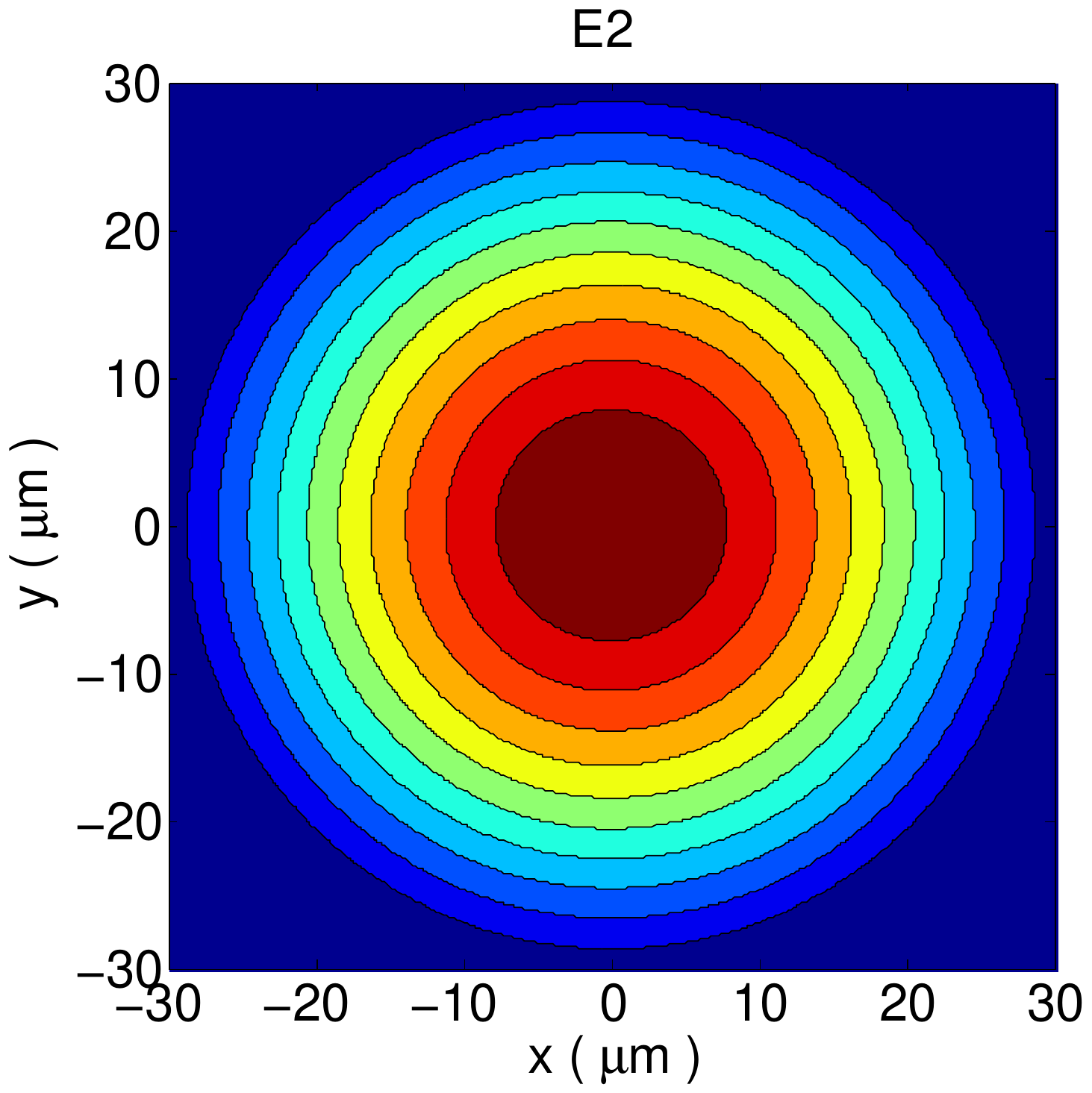}
\label{E2}
}
\subfigure{c)
\includegraphics[width=1.4 in]{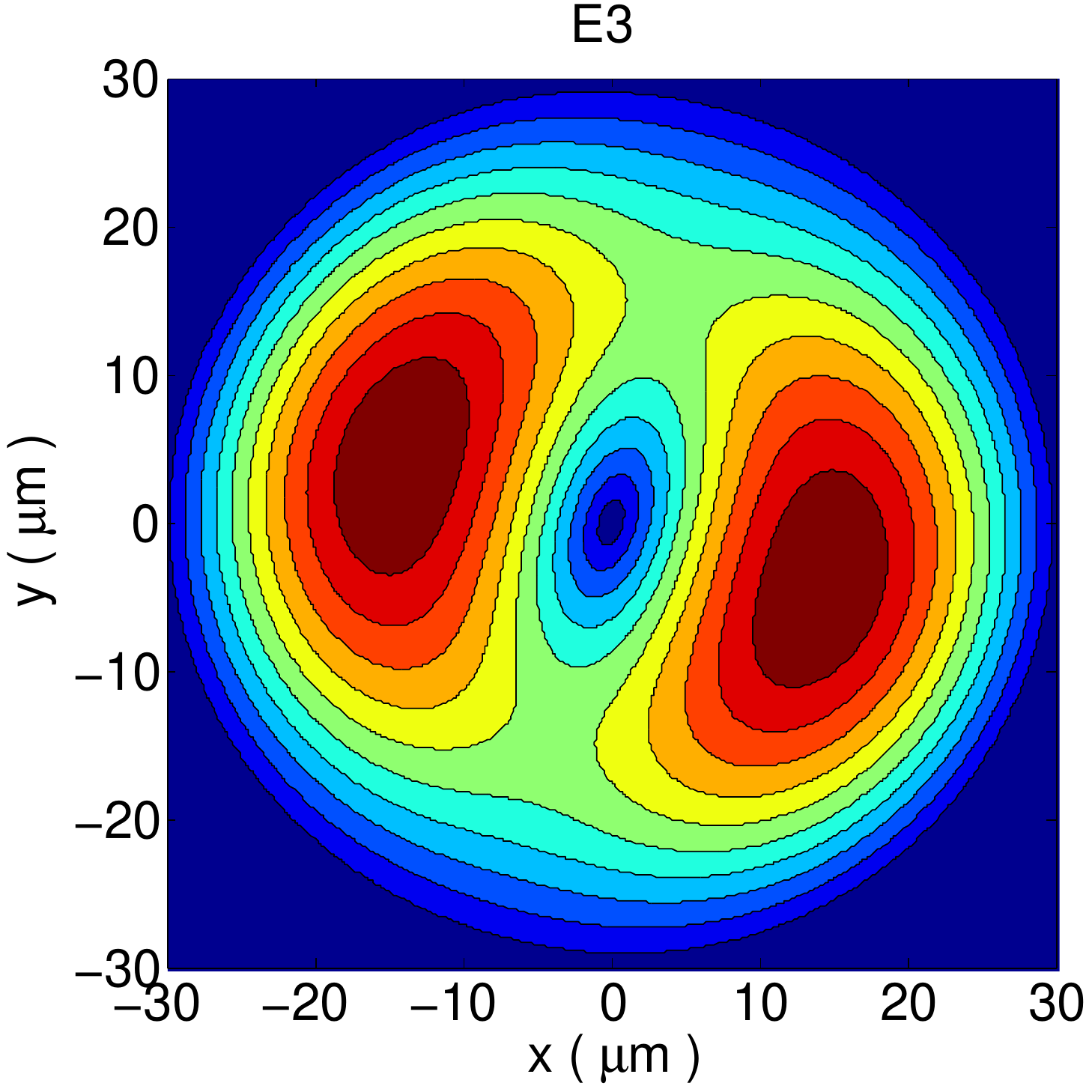}
\label{E3}
} 
\subfigure{d)
\includegraphics[width=1.4 in]{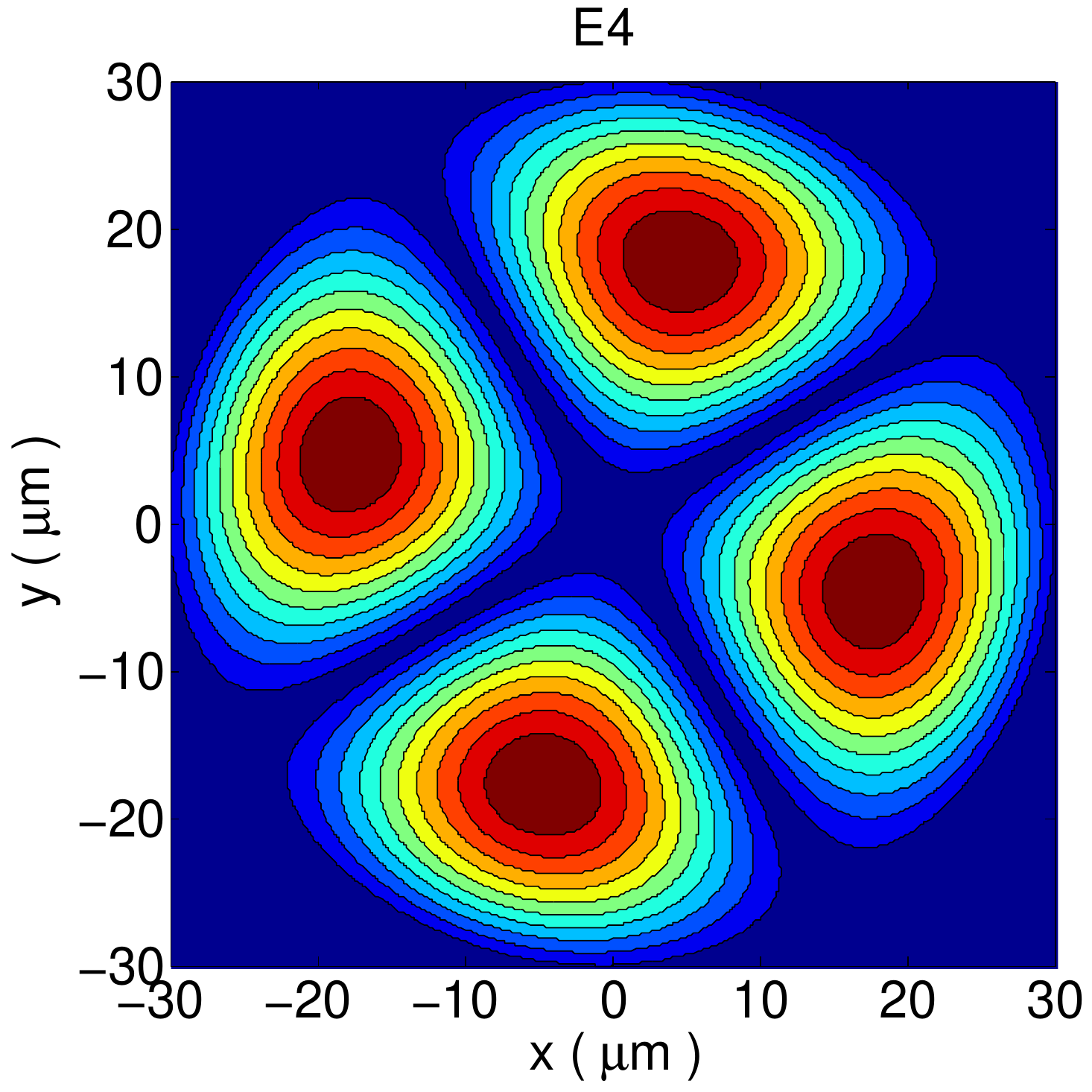}
\label{E4}
} 
\subfigure{e)
\includegraphics[width=1.4 in]{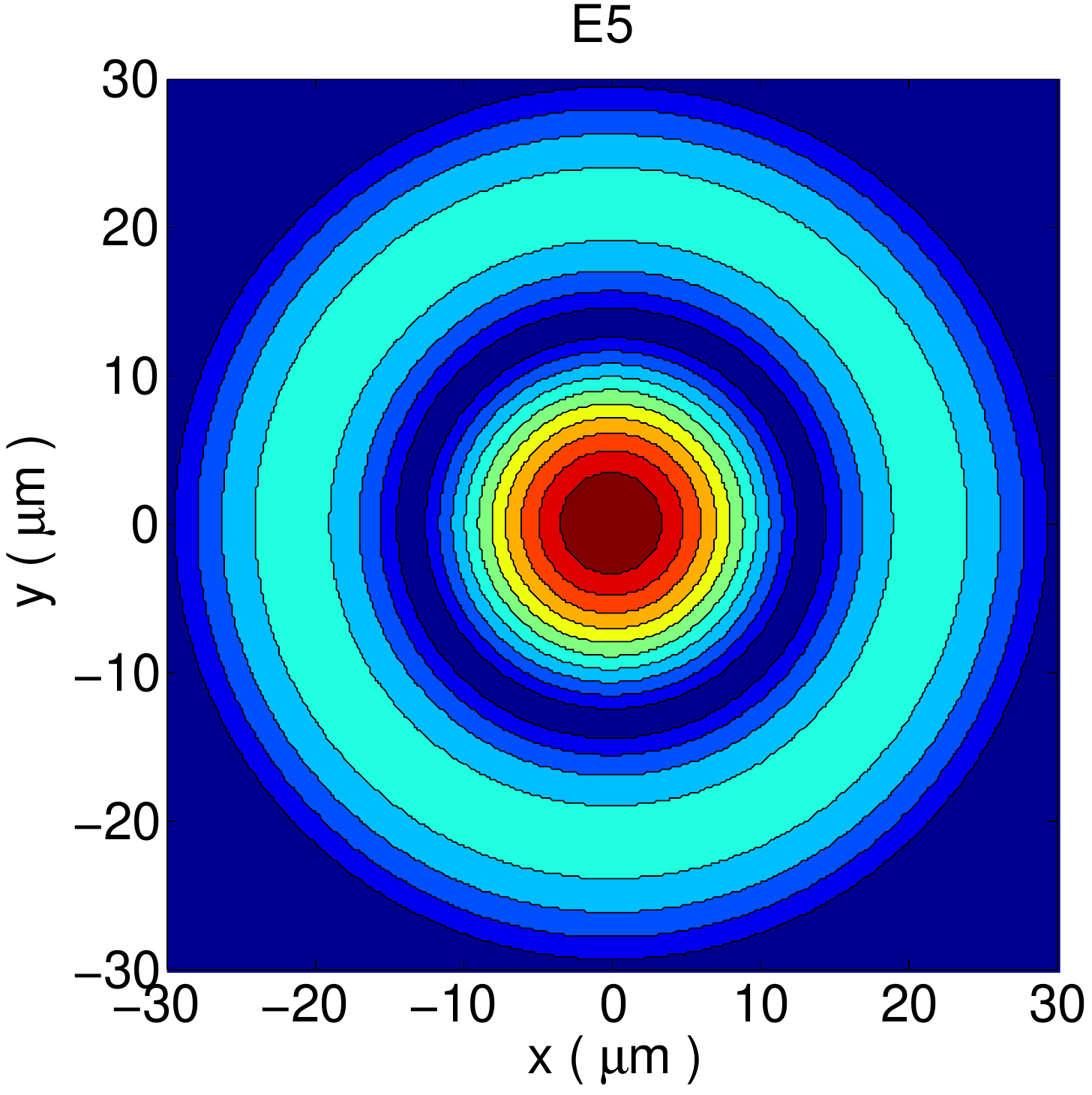}
\label{E5}
}
\subfigure{f)
\includegraphics[width=1.4 in]{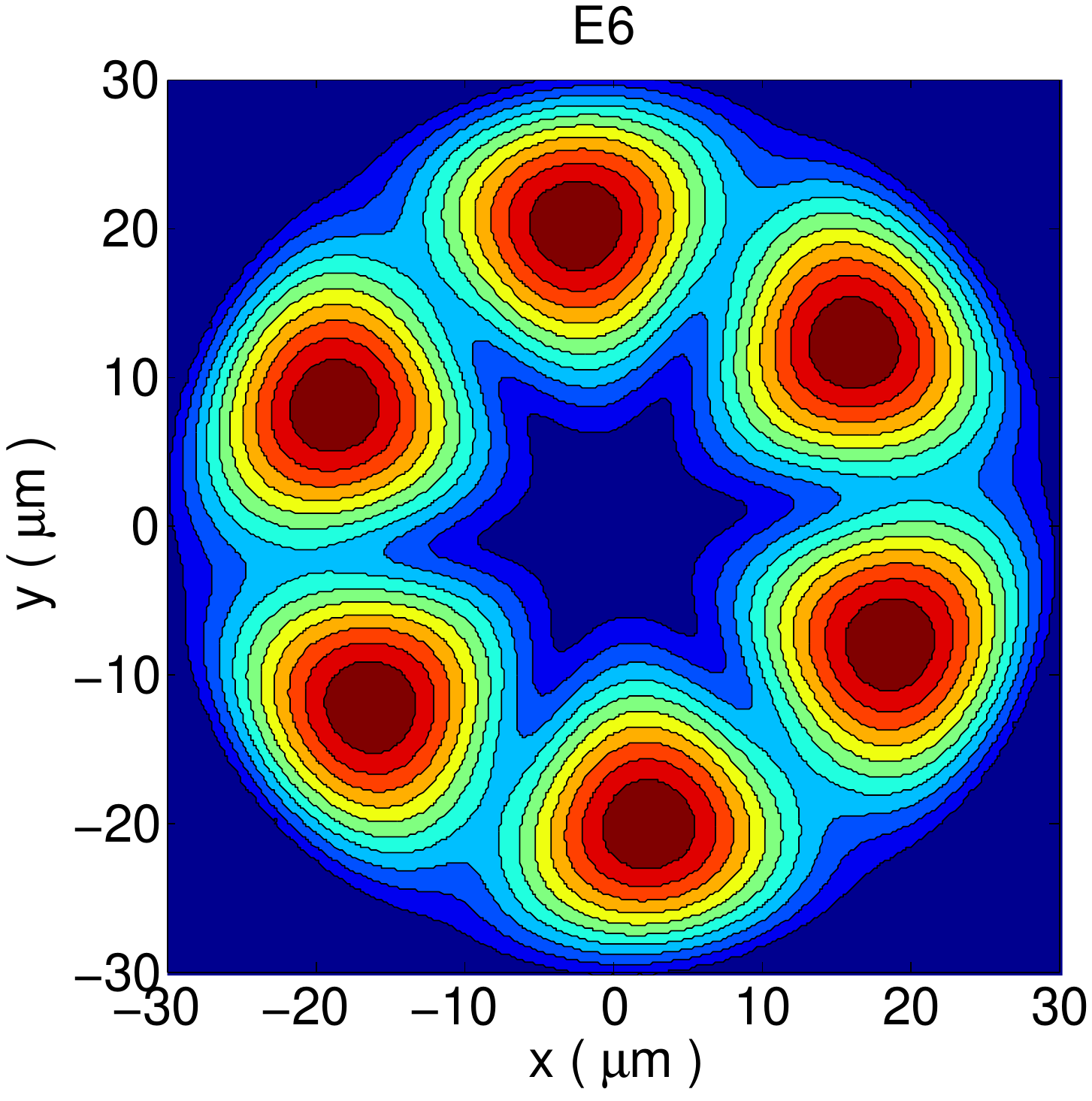}
\label{E6}
}
\caption{Graphic of $E_i(x,y)$. \subref{E1} LP01 mode $E_1$ with $\Delta x= \SI{15}{\micro\meter}$ and $\Delta y= \SI{10}{\micro\meter}$. \subref{E2} LP01 mode $E_2$. \subref{E3} LP11 mode $E_3$. \subref{E4} LP21 mode $E_4$. \subref{E5} LP02 mode $E_5$. \subref{E6} LP31 mode $E_6$.}
\end{figure}

\begin{figure} 
\centering
\includegraphics[width=2.6 in]{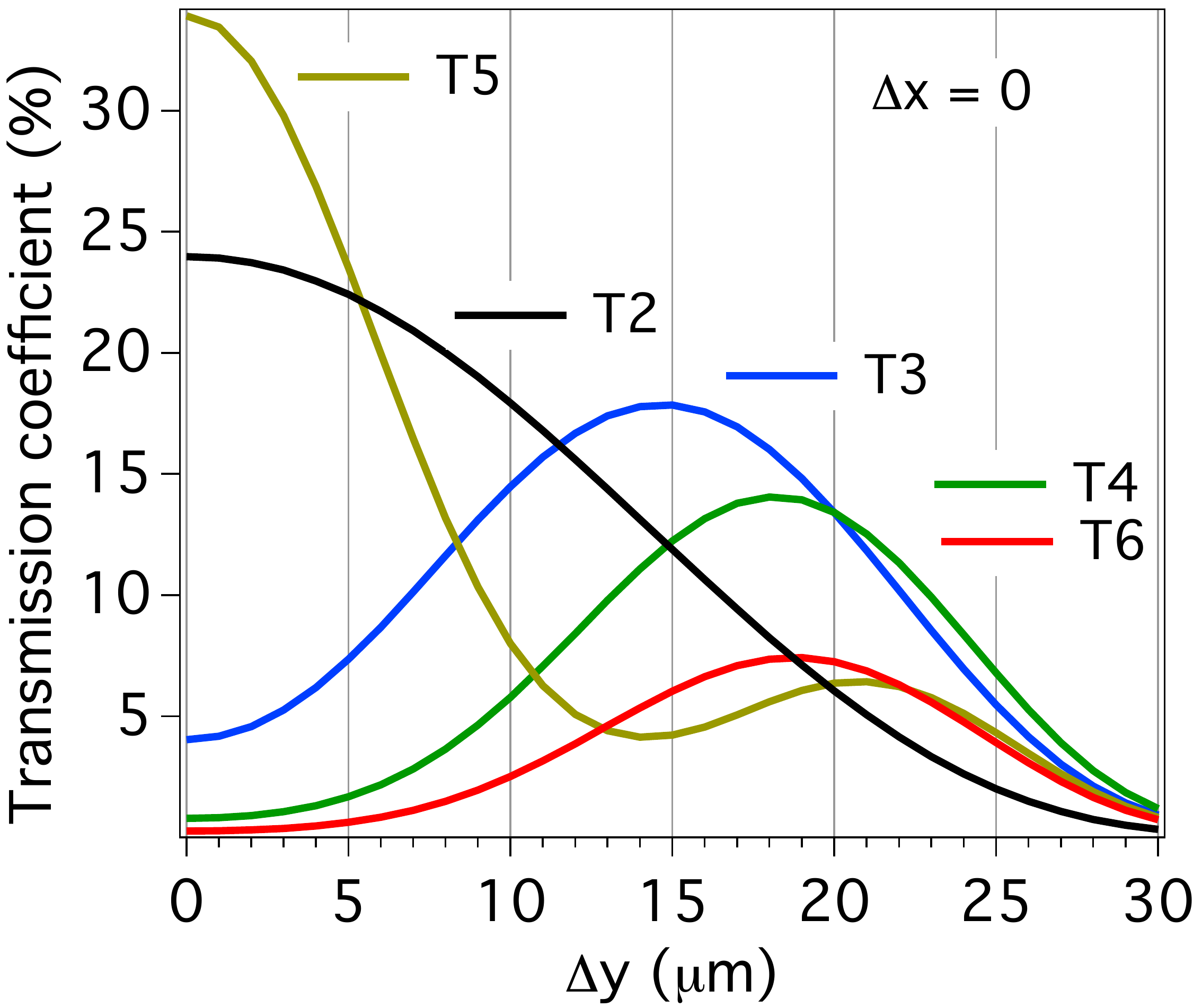}
\caption{Transmission $T_i$ as function of the transverse misalignment $\Delta y$ with $\Delta x = 0$.}
\label{transmissoes3}
\end{figure}

\section{Results}

\subsection{Transmission coefficients}

The computed modes on the SMF and MMF are displayed on Figs.~\ref{E1} to \ref{E6}. These can be identified as linearly polarized (LP) optical modes \cite{Eidam2011,Nicholson2012}: $E_1$ and $E_2$ are LP01, $E_3$ is LP11, $E_4$ is LP21, $E_5$ is LP02 and $E_6$ is LP31. The transverse misalignment was created by shifting the mode $E_1$ and letting the MMF modes be centered at the origin of the $xy$ plane. For example, the mode $E_1$ plotted on Fig.~\ref{E1} was displaced $\Delta x=\SI{15}{\micro\meter}$ and $\Delta y=\SI{10}{\micro\meter}$. Fig.~\ref{transmissoes3} shows the transmission coefficients as function of $\Delta y$ using~\eqref{transmisaofundamental} with $\Delta x = 0$. It can be seen how the transmission is maximum only when the maxima of the two modes match. We can clearly see that modes $E_2$ and $E_5$ have maximum transmission when the two fibers are aligned. Transmission to the fundamental mode $T_2$ is the second highest and $T_4$ has a smaller maximum which is almost the same intensity as the maximum of $T_6$. All these results show the relevance of the excited modes in the transmitted signal.

\begin{figure}
\centering
\subfigure{a)
\includegraphics[width=1.4 in]{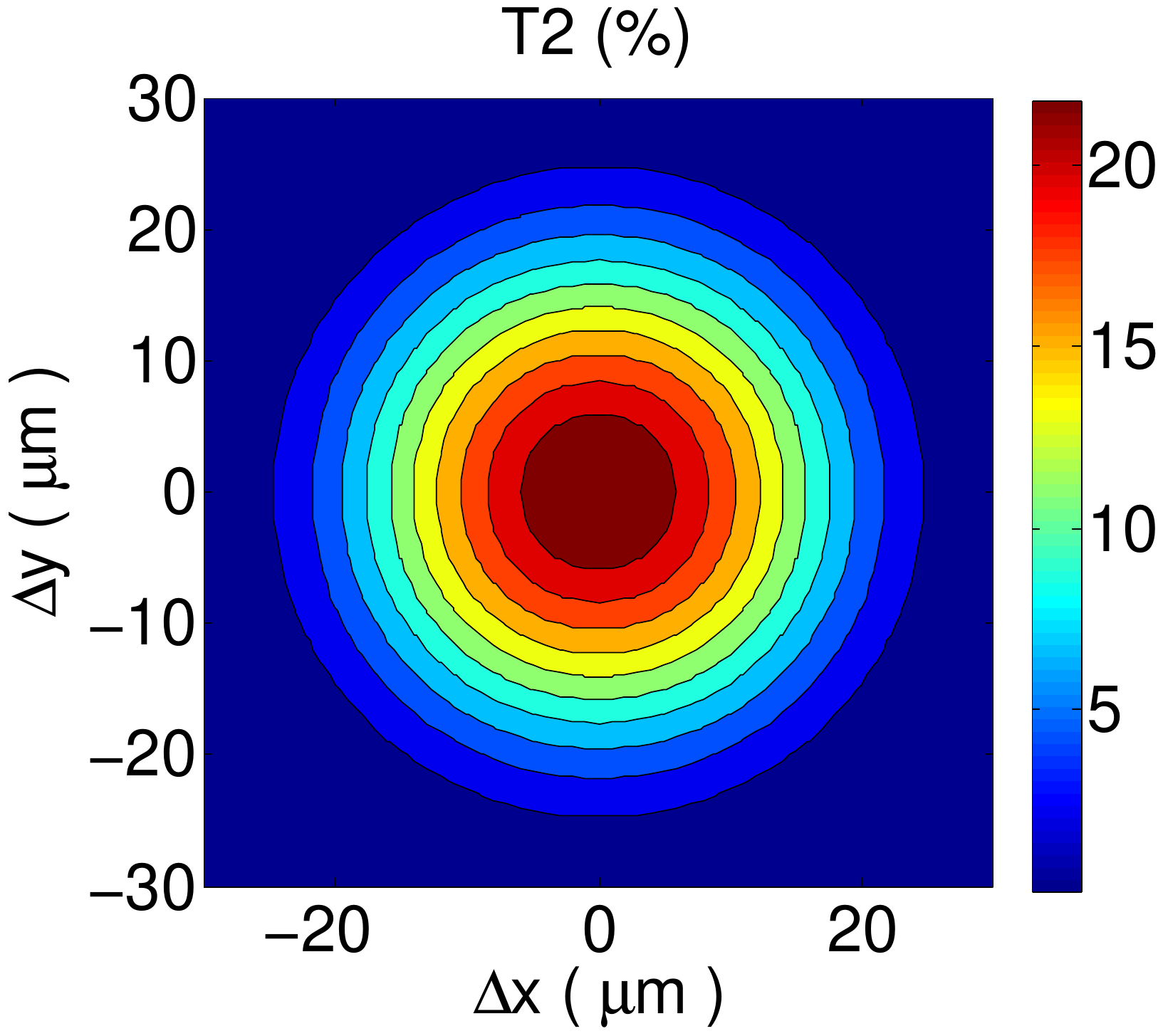}
\label{T12fig}
}
\subfigure{b)
\includegraphics[width=1.4 in]{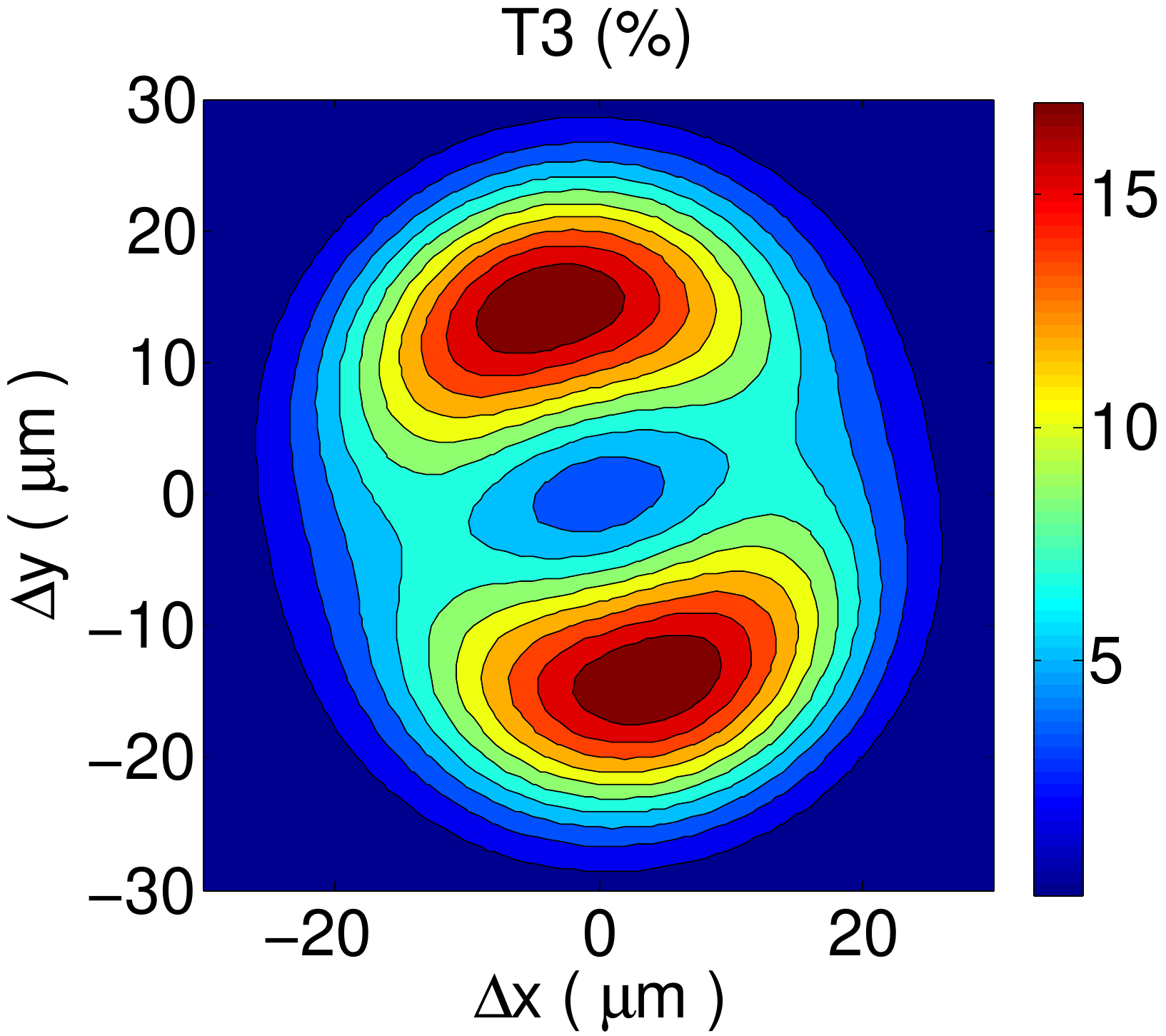}
\label{T13fig}
}
\subfigure{c)
\includegraphics[width=1.4 in]{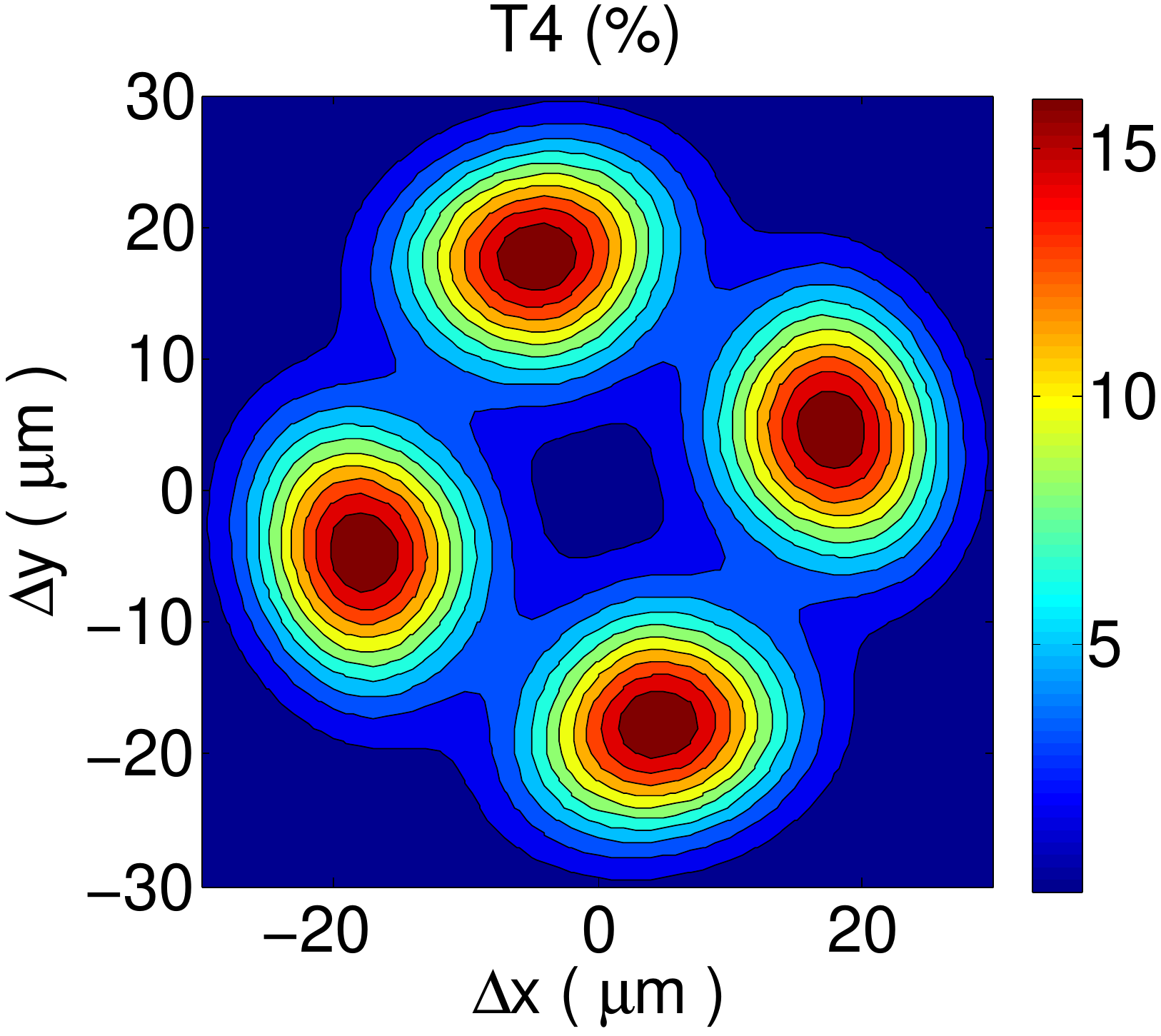}
\label{T14fig}
} 
\subfigure{d)
\includegraphics[width=1.4 in]{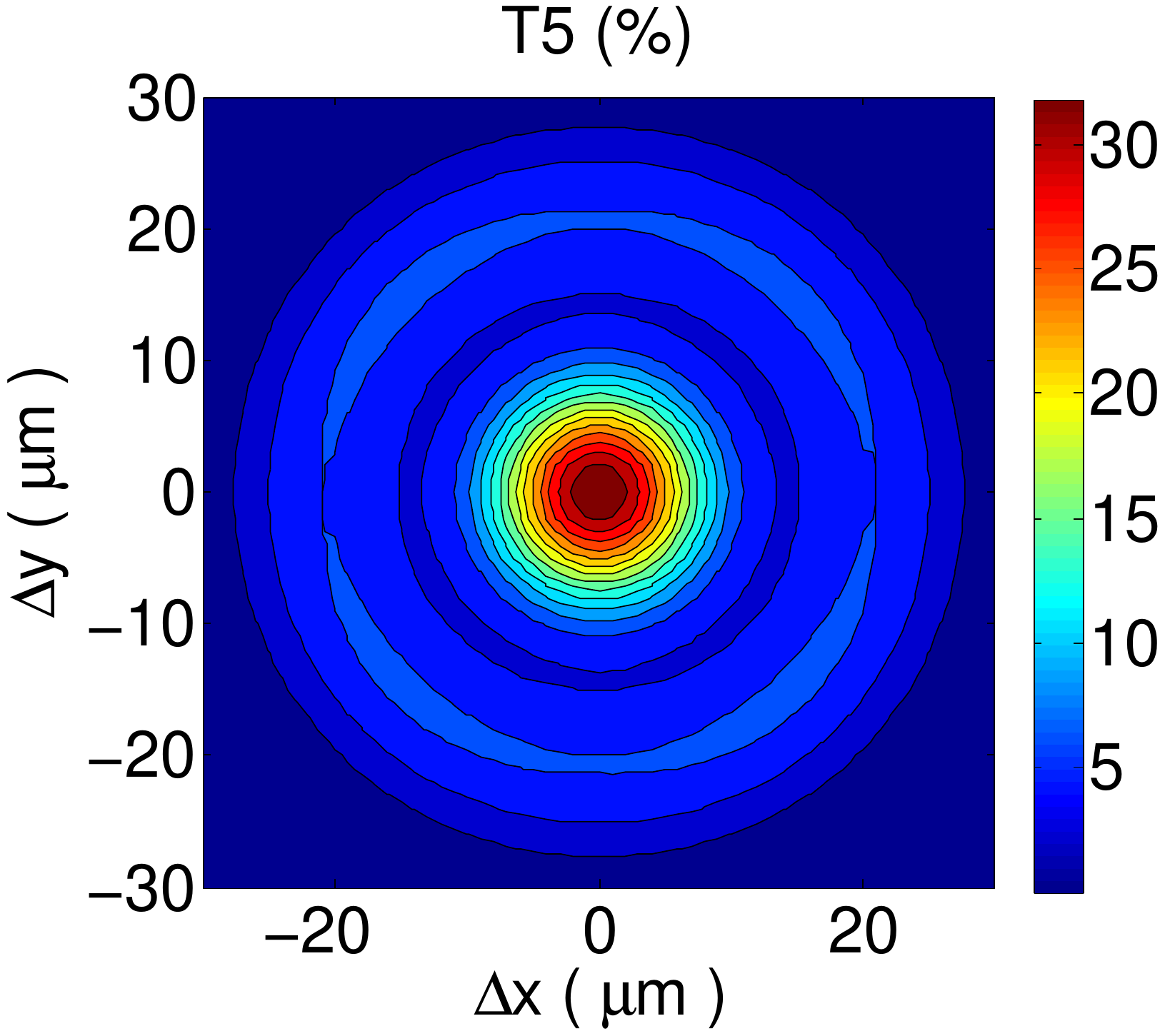}
\label{T15fig}
}
\subfigure{e)
\includegraphics[width=1.4 in]{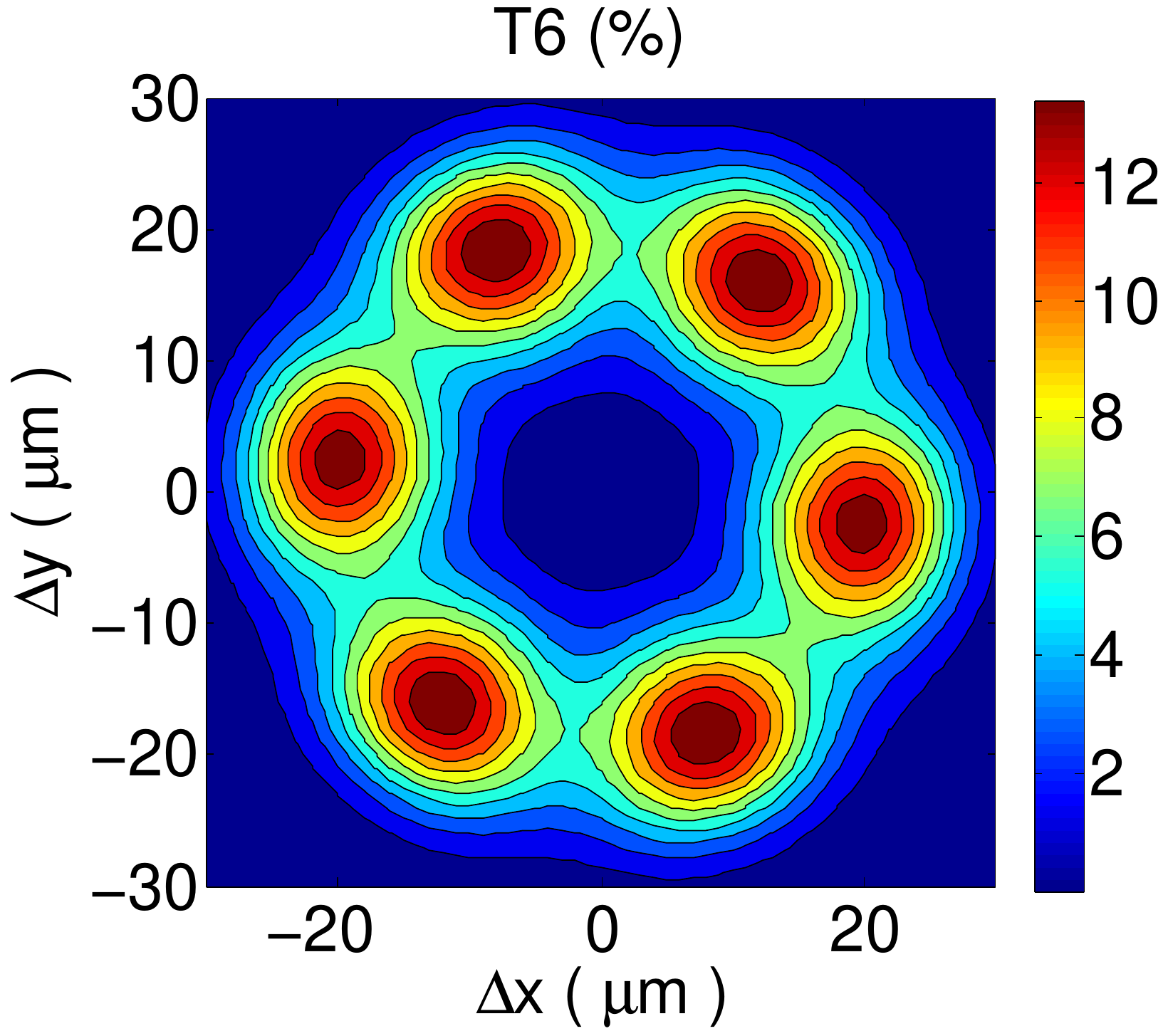}
\label{T16fig}
}
\subfigure{f)
\includegraphics[width=1.4 in]{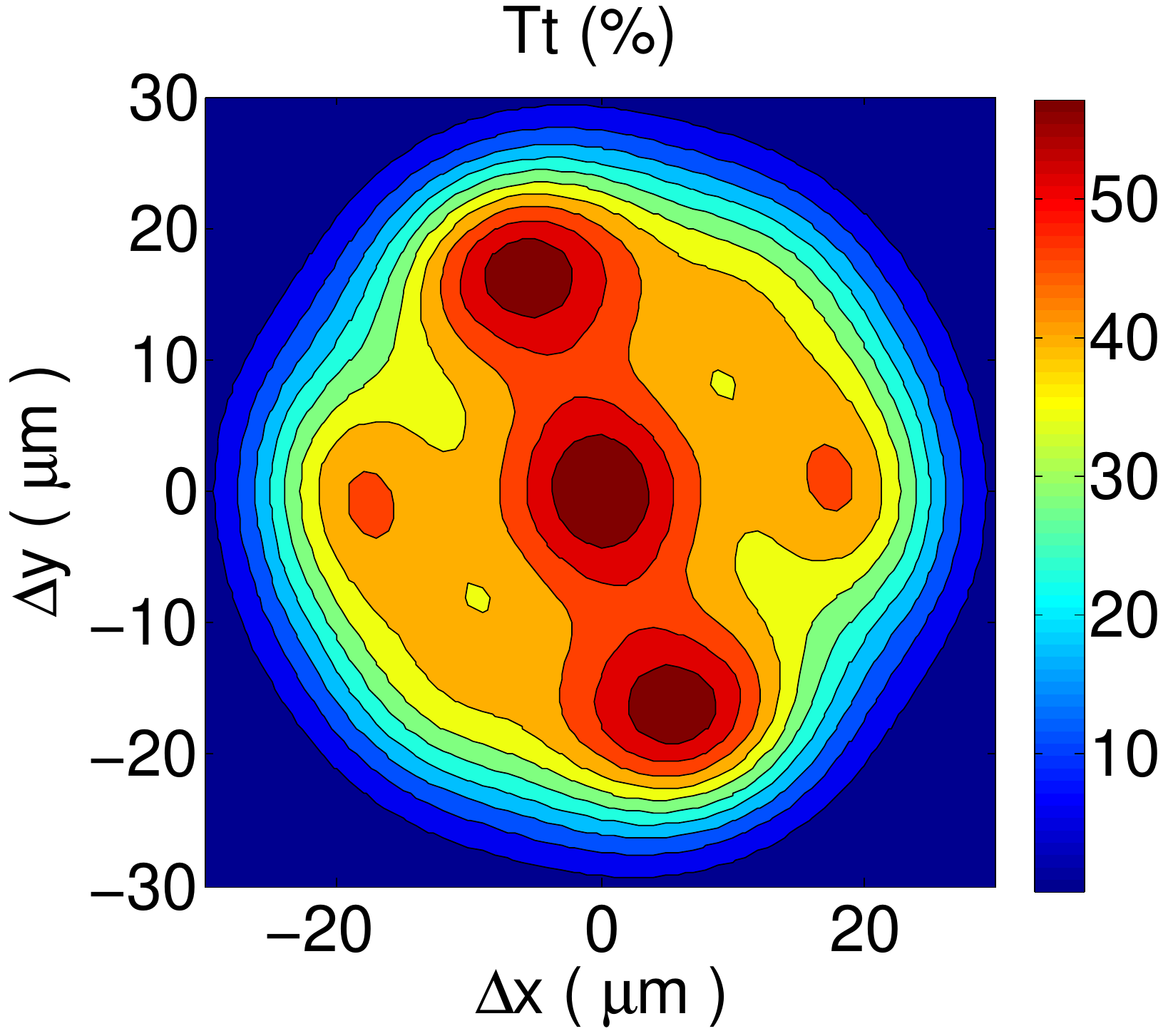}
\label{Ttotal}
}
\caption{Graphic of $T_i(\Delta x,\Delta y)$. \subref{T12fig} $T_2$ \subref{T13fig} $T_3$ \subref{T14fig} $T_4$ \subref{T15fig} $T_5$ \subref{T16fig} $T_6$. \subref{Ttotal} Total transmission $T_t$ (eq. \ref{transmissaototal}).}
\end{figure}

The computed transmission coefficients considering an arbitrary displacement $T_i(\Delta x, \Delta y)$ are displayed on the Figs. \ref{T12fig} to \ref{T16fig}. For each displacement $(\Delta x, \Delta y)$, the mode $E_1$ was shifted accordingly and then the integral $T_i$ (Eq. \ref{transmisaofundamental}) was computed. The scale on the color bar of Fig. \ref{T15fig} is the highest, in agreement with Fig. \ref{transmissoes3}. We can also see that $T_3$ has a maximum in the $\Delta y$ axis while $T_6$ has it in the $\Delta x$ axis, showing the lack of circular symmetry for the displacement, feature better seen on the total transmission coefficient $T_t$ (displayed on Fig. \ref{Ttotal}). It has a primary maximum in the center and two other secondary maxima at $(\Delta x, \Delta y) = (\pm 6, \mp 16)$ with 97 \% of the intensity of the primary one.

\subsection{Transmitted power of the SMS device}

\begin{figure}
\centering
\subfigure{a)
\includegraphics[width=2.1 in]{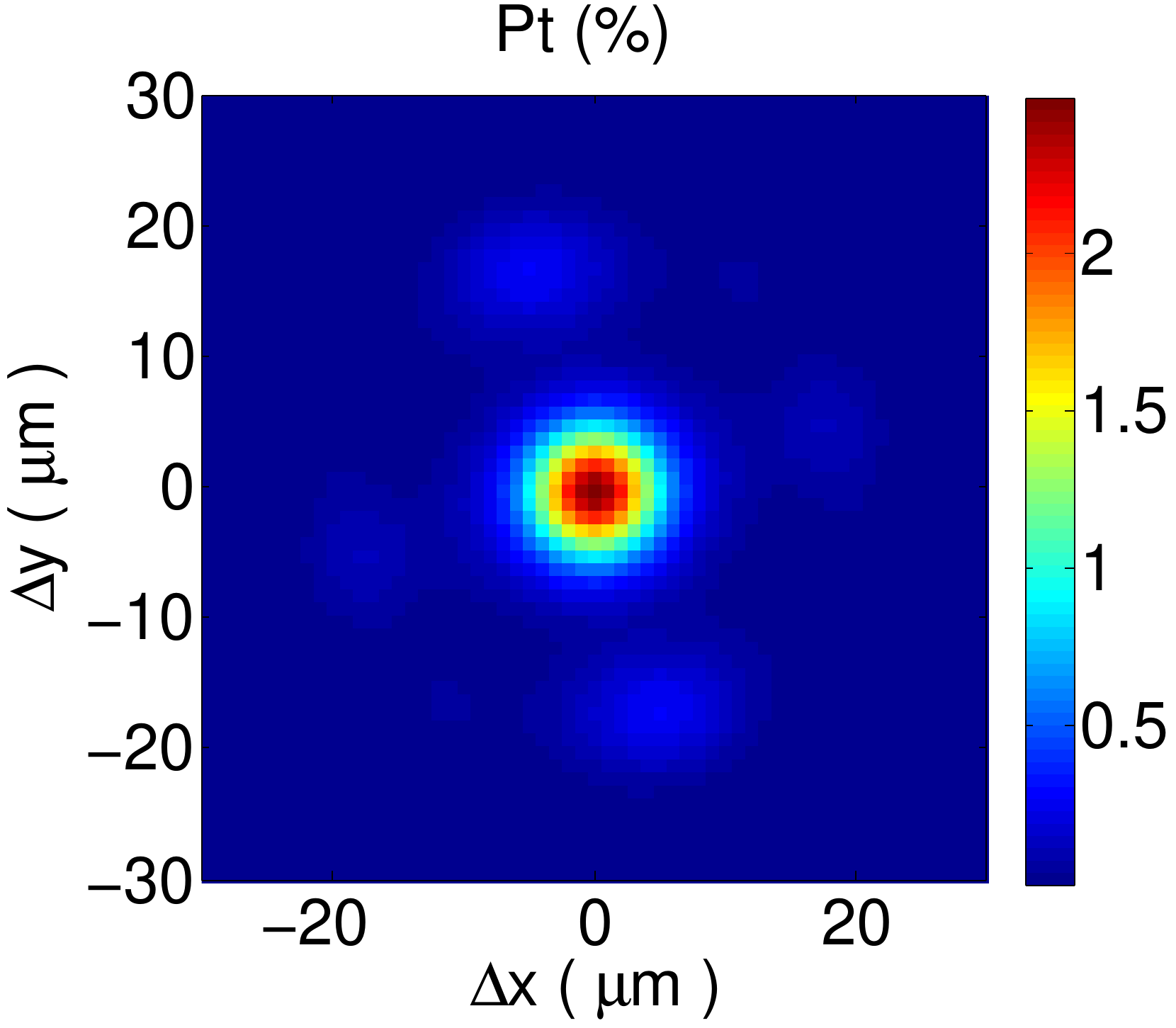}
\label{Ptotal2}
}
\subfigure{b)
\includegraphics[width=2.1 in]{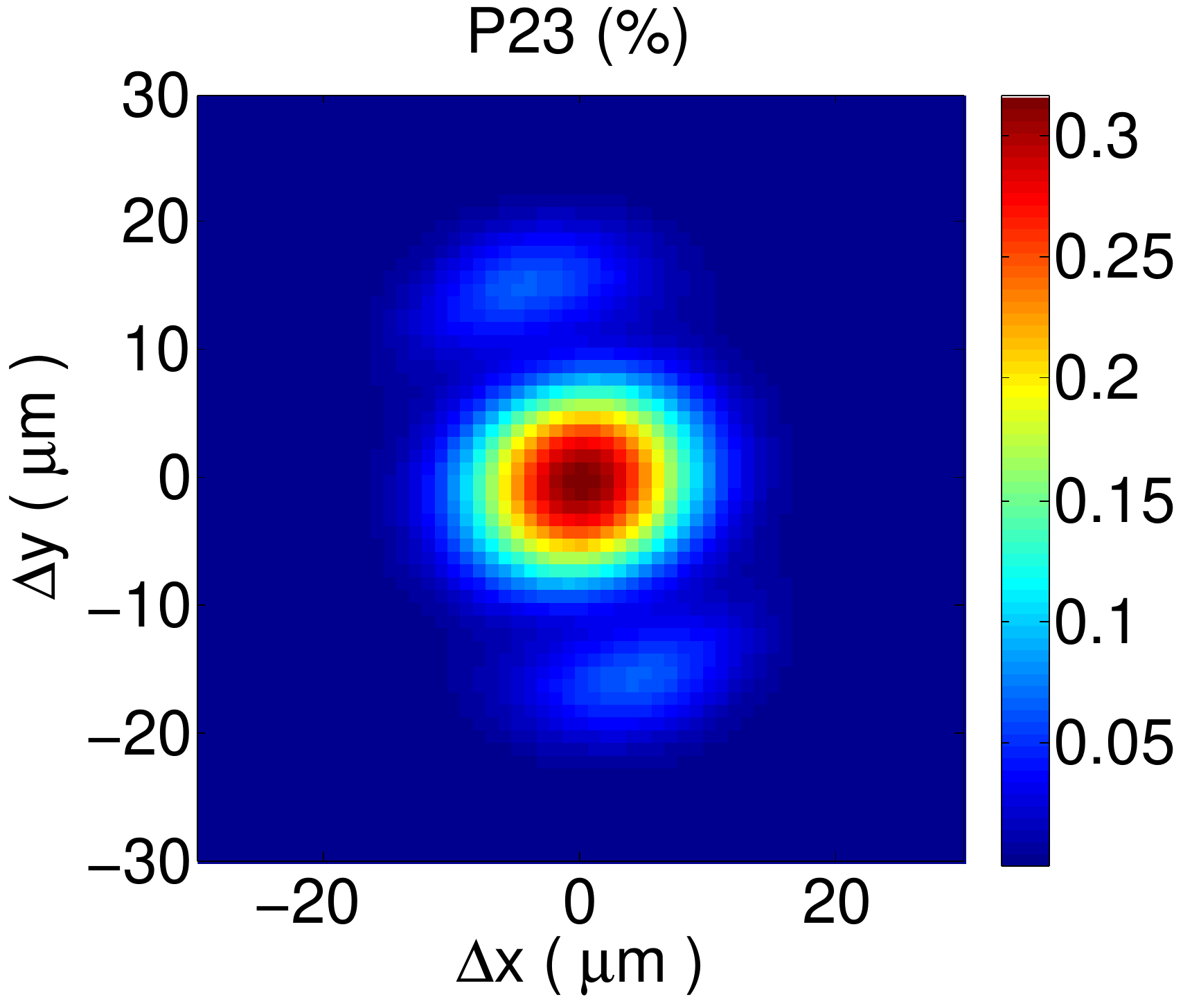}
\label{P23}
}
\caption{Graphic of the transmitted power over the SMS sensor for an arbitrary displacement at room temperature $T_0 =24^{\circ}C$. \subref{Ptotal2}. Power $P_t$ (Eq. \ref{powertotal}). \subref{P23} Power $P_{23}$ (Eq. \ref{powerP23}).}
\end{figure}

In Fig. \ref{Ptotal2} it is shown the graphic of SMS power $P_t$ as in~\eqref{powertotal}, where the principal maximum is in the center (no displacement), two secondary maxima at $(\Delta x, \Delta y) = (\pm 5, \mp 17)$ and other two tertiary maxima at $(\Delta x, \Delta y) = (\pm 17, \pm 5)$. As before, for each displacement $(\Delta x, \Delta y)$ one integral was performed. The intensities of the secondary and tertiary maximum are 11 and 4.8\,\% of the primary maximum intensity. 

In the case of the SMS sensor there are two interfaces between the two fibers which reduces the intensity of the misaligned maxima. The graphic of $P_{23}$ (Eq. \eqref{powerP23}) is shown on the Fig.~\ref{P23}. A primary maximum is in the center and two secondary ones are at $(\Delta x, \Delta y) = (\pm 5, \mp 15)$ and have 20\,\% of primary maximum intensity. This maximum intensity is one order or magnitude lower than the one in Fig. \ref{Ptotal2}. These results show that although the excited modes $E_4$, $E_5$ and $E_6$ were not significant enough to create maxima for the misaligned case $(\Delta x,\Delta y) \neq (0,0)$, they contributed to the maximum intensity in the aligned case $\Delta x =\Delta y =0$.

\begin{figure} 
\centering
\includegraphics[width=2.6 in]{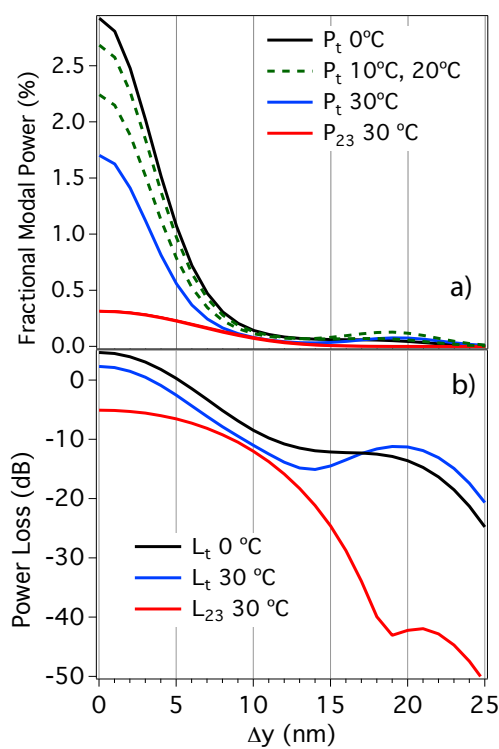}
\caption{Transmission over the SMS sensor vs $\Delta y$ for different temperatures. a) Transmitted power $P_t$ (eq. \ref{powertotal}) and $P_{23}$ (Eq. \ref{powerP23}). b) Power loss $L_t$ and $L_{23}$ (Eqs. \ref{dfgbnmzaqwe}).}
\label{fmpowerloss}
\end{figure}

The effect of the temperature is shown on the Fig. \ref{fmpowerloss}(a) with the graphic of $P_t$ and $P_{23}$ vs. $\Delta y$ for the temperatures 0, 10, 20, and \SI{30}{\celsius}. For $\Delta y>\SI{15}{\micro\meter}$, $P_t$ and $P_{23}$ are negligible, which indicates that the transmitted power is significant only in the aligned case. In this case ($\Delta x =\Delta y =0$), a lower temperature implies in a larger contribution of the excited modes $E_4$, $E_5$ and $E_6$. Fig. \ref{fmpowerloss}(b) shows the graphics of the power loss $L$ as function of the temperature. The power loss $L_{23}$ (considering only the modes $E_2$ and $E_3$) was rather insensitive to the temperature, reason why only the one at \SI{30}{\celsius} is shown. For all $\Delta y$ we have $L_t>L_{23}$, specially for $\Delta y>\SI{15}{\micro\meter}$, showing the importance of the excited modes in the lower temperature case. Again, the power loss is significant only in the aligned case.

\section{Conclusions}

As conclusions, we have studied the transmission through an SMS device by numerically evaluating the superposition integrals of the electromagnetic modes. 
We computed the transmission considering five modes on the multimode fiber as function of the transverse misalignment between the fibers. We observed that when modes without circular symmetry are considered there can be maxima in the transmission with the two fibers misaligned. We also included the effect of the temperature and we observed that the transmitted power is significant only in the aligned case and that the lower temperature implies in a larger contribution of the excited modes $E_4$, $E_5$ and $E_6$.

\section{Acknowledgments}

The authors thank brazilian agency CNPq. This work was partially supported by Finep/Funttel Grant No. 01.14.0231.00, under the Radiocommunication Reference Center (Centro de Refer\^{e}ncia em Radicomunica\c{c}\~{o}es - CRR) project of the National Institute of Telecommunications (Instituto Nacional de Telecomunica\c{c}\~{o}es - Inatel), Brazil.

\end{document}